\documentclass[12pt]{article}
\usepackage{amsmath,epsfig,bm,mathrsfs}

\numberwithin{equation}{section}
\newcommand\eq[1]{(\ref{#1})}

\renewcommand\d{\partial}
\newcommand\<{\langle}
\renewcommand\>{\rangle}

\newcommand\lb{\label}
\newcommand\Tr{{\rm Tr }}
\newcommand\JV{\mathscr{V}}
\newcommand\JA{\mathscr{A}}
\newcommand\JL{\mathscr{L}}
\newcommand\JR{\mathscr{R}}
\newcommand\dg{|g|}
\newcommand\emnlrs{\epsilon^{{\hat\mu}{\hat\nu}{\hat\lambda}{\hat\rho}{\hat\sigma}}}

\begin{document}
\title{QCD and dimensional deconstruction}
\author{D.~T.~Son$^1$ and M.~A.~Stephanov$^2$\vspace{8pt}\\
{\em\small $^1$Institute for Nuclear Theory, University of Washington, 
Seattle, WA 98195}\\
{\em\small $^2$Department of Physics, University of Illinois, Chicago, IL 
60607-7059}\\
{\em\small $^3$RIKEN-BNL Research Center, Brookhaven National Laboratory,
Upton, NY 11973}}
\date{April 2003}
\maketitle

\begin{abstract}
Motivated by phenomenological models of hidden local symmetries
and the ideas of dimensional deconstruction and gauge/gravity duality, we
consider the model of an ``open moose.'' Such a model has
a large number $K$ of hidden gauge groups as well as a global chiral
symmetry.  In the continuum limit $K\to\infty$ the model
becomes a 4+1 dimensional theory of a gauge field propagating 
in a dilaton background and
an external space-time metric with two boundaries. We show that
the model reproduces several well known phenomenological and
theoretical aspects of low-energy hadron dynamics, 
such as vector meson dominance.
We derive the general
formulas for the mass spectrum, the decay constants of the pion and
vector mesons, and the couplings between mesons.  We then consider
two simple realizations, one with a flat metric and another with a
``cosh'' metric interpolating between two AdS boundaries.  For the
pion form factor, the single pole 
$\rho$-meson dominance is exact in the latter case
and approximate in the former case.  We discover that an AdS/CFT-like
prescription emerges in the computation of current-current correlators.
We speculate on the role of the model in the theory dual to QCD.
\end{abstract}

\newpage

\section{Introduction}

Vector mesons ($\rho(770)$, $\omega(782)$, etc.) play a significant role
in hadronic physics.  Their interactions, though not constrained by
low-energy theorems, apparently follow the broad pattern of vector meson
dominance (VMD)~\cite{Sakurai}.  There have been numerous efforts to
incorporate vector mesons into field-theoretical frameworks.
Historically, the Yang-Mills theory was discovered in an early attempt
to treat the $\rho$ meson~\cite{YM}.  More recently, interesting
schemes based on ``hidden local symmetries'' (HLS)
were developed by Bando {\em et
al.}~\cite{HLS,HLSa1,HLSreview}.  
In the original model~\cite{HLS}, the $\rho$ meson is the
boson of a spontaneously broken gauge group.  The model has been
extended to two hidden gauge groups~\cite{HLSa1}; then it also
incorporates the lowest axial vector meson $a_1(1260)$.  With suitable
parameters, these models can be quite successful phenomenologically,
although they cannot be systematically derived from QCD (except in the
limit of very light $\rho$, if such a limit could be
reached~\cite{Georgi}).

In this paper we explore theories with very large, and even infinite
number $K$ of hidden local symmetries.  Our motivation is twofold.
First and most straightforwardly,
there are excited states in the vector and axial vector
channels ($\rho(1450)$, $a_1(1640)$, $\rho(1700)$, etc.~\cite{PDG}),
which must become narrow resonances in the limit of large number of colors
$N_c$.  It is tempting to treat them as gauge bosons of
additional broken gauge groups.\footnote{To our knowledge, the
  earliest attempt to interpret the tower of $\rho$, $\rho'$, etc.
as a ``chain structure'' was made in Ref.~\cite{HalpernSiegel}.}

The second motivation comes from
recent theoretical developments.  Many strongly coupled gauge theories
are found to have a dual description in terms of theories with gravity
in higher dimensions~\cite{Maldacena,GKP,Witten,AdSreview}.  
It was suggested
that the string theory dual to large-$N_c$ QCD must have strings
propagating in five dimensions, in which the fifth dimension has the
physical meaning of the energy scale~\cite{Polyakov-cave}.  In the
framework of field theory, the fifth dimension can be
``deconstructed'' in models with a large number of gauge
fields~\cite{deconstr,HPW}.

We discovered that the continuum limit $K\to\infty$ can lead to results
that qualitatively, and in many cases even quantitatively, agree with
phenomenology. Most remarkably,
the vector meson dominance, which in the HLS theories required a
tuning of parameters, becomes a natural consequence of the
$K\to\infty$ limit. Another advantage of  the limit
$K\to\infty$ is the possibility of matching to the asymptotic
behavior of the current-current correlator known from perturbative QCD.

As anticipated, a natural interpretation of
this limit is a discretization, or deconstruction, of a
5-dimensional gauge theory.  Further, to our amusement, in the
calculation of current-current correlators we found a relation very
similar to the one employed in the AdS/CFT correspondence: the
current-current correlator in 4d theory is expressed in terms of the
variations of the classical 5d action with respect to the boundary
values of the bulk gauge fields on the 4d boundaries.

We limit our discussion to the isospin-1 sector of QCD.
It is straightforward to extend the discussion to the isospin-0
sector ($\eta$, $\omega$, and $f_1$ mesons).
The detailed treatment of the $U(1)_A$ problem, 
chiral anomaly, Wess-Zumino-Witten term, and 
 baryons is deferred to future work.

The paper is organized as follows.  In section~\ref{sec:model} we
describe the open moose model.  In section~\ref{sec:observables} we
compute different physical observables: the vector meson mass
spectrum, the decay constants of the pion and the vector mesons, the
coupling between the vector mesons and the pions, and the pion
electromagnetic form factor.  We also check the validity of Weinberg's
spectral sum rules, and discover that the limit $K\to\infty$
automatically leads to exact VMD for the pion formfactor.

In section~\ref{sec:continuum} we take the limit
of infinite number of the hidden groups $K\to\infty$.  We show that
the theory can be understood as a 5d Yang-Mills theory in an external
metric and dilaton background.  We establish an AdS/CFT-type
prescription for calculating the current-current correlators.  We
consider two concrete realizations of the open moose in
section~\ref{sec:examples}. 
We find that a ``cosh'' background metric interpolating between two
AdS boundaries leads to correct asymptotic behavior of the
current-current correlator. This allows us to establish a relationship
between hadron parameters such as $f_\pi$, $m_\rho$, and the QCD
parameter $N_c$.
 In section~\ref{sec:baryon} we show that
the instanton, which is a quasiparticle in $4+1$ dimensions, becomes a
Skyrmion upon reduction to 4d, and thus describes the baryon.
Section~\ref{sec:concl} contains concluding remarks.

\section{The open moose}
\label{sec:model}

The model under consideration is described by the 
following Lagrangian%
\footnote{We are using the 
usual 3+1 Minkowski metric $\eta_{\mu\nu}={\rm diag}(1,-1,-1,-1)$,
but  write all indices as lower indices
for simplicity, unless it could lead to a confusion.}
\begin{equation}
{\cal L} = \sum_{k=1}^{K+1} f_k^2\Tr |D_\mu \Sigma^k |^2
-
\sum_{k=1}^{K} \frac12 \Tr \left(F_{\mu\nu}^k\right)^2.
\lb{L}
\end{equation}
The covariant derivatives are defined as
\begin{subequations}
\begin{eqnarray}
  D_{\mu}\Sigma^1 &=& \partial_\mu\Sigma^1 
  + i\Sigma^1(gA_\mu)^1\label{dsigma1},\\  
  D_{\mu}\Sigma^k &=& \partial_\mu\Sigma^k - i(gA_\mu)^{k-1}\Sigma^k 
  + i\Sigma^k(gA_\mu)^{k}\label{dsigma},\\
  D_\mu \Sigma^{K+1} &=& \d_\mu\Sigma^{K+1} -i(gA_\mu)^K\Sigma^{K+1}.
  \label{dsigma3}
\end{eqnarray}
\end{subequations}
A shorthand notation is used for the product of the gauge field
$A_\mu=A_\mu^a\tau^a/2$ and its coupling constant: $g_k A_\mu^k
\equiv (gA_\mu)^k$.  If we assume $A_\mu^0=A_\mu^{K+1}=0$, then 
Eqs.~(\ref{dsigma1}) and (\ref{dsigma3})
become special cases of
Eq.~(\ref{dsigma})
for $k=1$ and $k=K+1$.

The model contains $K+1$ nonlinear sigma model fields $\Sigma^k\in
\textrm{SU(2)}$ (or, in general, $\textrm{SU}(N_f)$), interacting via
$K$ ``hidden'' gauge bosons $A^k_\mu$.  
The model has a chiral $\textrm{SU(2)}\times\textrm{SU(2)}$ symmetry and
an $\textrm{SU(2)}^K$ local symmetry:
\begin{equation}
\begin{split}\lb{usigmau}
  \Sigma^1 &\to L\Sigma^1 U_1^\dagger(x)\,, \\
  \Sigma^k &\to U_{k-1}(x)\Sigma^k U_k^\dagger(x)\,,\qquad k=2,\,3,\ldots\,,\\
  \Sigma^{K+1} &\to U_K(x) \Sigma^K R^\dagger.
\end{split}
\end{equation}
In particular, the product 
\begin{equation}\label{Sigmaproduct}
  \Sigma=\Sigma^1\Sigma^2\cdots\Sigma^{K+1}
\end{equation}
is the pion field, which can be seen from its transformation
properties,
\begin{equation}
  \Sigma \to L\Sigma R^\dagger.
\end{equation}

The parameters entering (\ref{L}) are $K+1$ decay constants $f_k$ and
$K$ gauge couplings $g_k$.  We shall assume they are invariant under a
reflection with respect to the middle of the chain,
\begin{equation}
  f_k = f_{K+2-k}\,,\qquad g_k = g_{K+1-k}\,,
\end{equation}
which ensures parity is a symmetry in the theory \eq{L}.

In the case $K=0$ the model reduces to the
chiral Lagrangian.  For $K=1$ it is the version of the hidden local
symmetry realized in the limit of very light $\rho$'s~\cite{Georgi}.
The model with $K=2$ and a particular choice of parameters
$f_1^2=f_3^2=2f_2^2$, $g_1=g_2$ has been considered in
Ref.~\cite{HLSa1}.  Graphically, the model can be represented by a
``theory-space'' diagram shown in Fig.~\ref{fig:K}.  Since this
diagram is the usual ``moose diagram'' cut open, we shall call the
model~(\ref{L}) the ``open moose'' theory.

\begin{figure}
\begin{center}
\epsfig{file=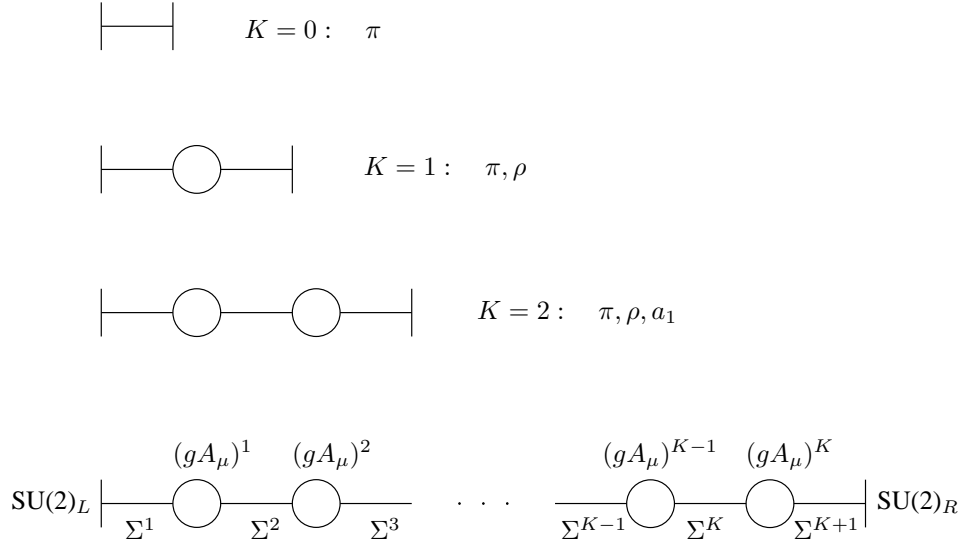}
\end{center}
\caption[]{Graphic representation of the Lagrangian \eq{L}. The examples
of low $K$ corresponding to previously considered theories are also shown.
$K=0$ represents Weinberg's nonlinear sigma model for pions;
$K=1$ represents a hidden symmetry Lagrangian
describing $\pi$ and $\rho$; and $K=2$ represents a description of 
$\pi$, $\rho$, and
$a_1$ \cite{HLS,HLSa1}.}
\lb{fig:K}
\end{figure}

Note that
\eq{L} is not the most general Lagrangian satisfying all the symmetries
and limited to lowest derivatives.  In fact, terms of the
following type are not forbidden:
\begin{equation}
  |\d_\mu(\Sigma^k\Sigma^{k+1}) - i(gA_\mu)^{k-1}(\Sigma^k\Sigma^{k+1})
  + i (\Sigma^k\Sigma^{k+1})(gA_\mu)^{k+1}|^2,
\end{equation}
as well as analogous expressions containing products of more than two
consecutive $\Sigma$'s.
In order to restrict the Lagrangian to the form (\ref{L}) an additional
condition of nearest neighbor locality in the  $k$ space
should be imposed. It is this condition that enables us later
to interpret this theory as a dimensionally deconstructed 
5d gauge theory in the limit $K\to\infty$.

\section{Physical observables}
\label{sec:observables}

In this section we derive expressions for physical observables,
 such as pion and vector meson decay constants, mass spectrum,
and pion-vector couplings in terms of the parameters of the moose
$f_k$ and $g_k$.

\subsection{Pion decay constant $f_\pi$}

For computations, the following gauge is the most convenient,
\begin{equation}\lb{SigmaPi}
\Sigma^k = \exp\left\{i\Pi\frac{f}{2f_k^2}\right\},
\quad \mbox{where}\quad \Pi = \pi^a \frac{\tau^a}2\,,
\end{equation}
where $f$ is a function of all $f_k$, which we shall specify in a
moment (in Eq.~\eq{f}).
The advantage of this gauge is that the pion field $\Pi$ does not
mix with other fields:
\begin{equation}
{\cal L}_{\rm mix} = \sum_{k=1}^{K+1} f\Tr\partial_\mu\Pi 
\left[(gA_\mu)^k - (gA_\mu)^{k-1}\right] = 0.
\end{equation}

The value of $f$ is fixed by requiring that the kinetic term
for $\pi^a$ is canonically normalized:
\begin{equation}
{\cal L}_{\pi^2} = \sum_{k=1}^{K+1} \frac{f^2}{4f_k^2} 
\Tr(\partial_\mu\Pi)^2 = \sum_{k=1}^{K+1} \frac{f^2}{4f_k^2} \, 
\frac12(\partial_\mu\pi^a)^2.
\end{equation}
Therefore
\begin{equation}
\frac4{f^2} = \sum_{k=1}^{K+1}\frac1{f_k^2}\,.
\lb{f}
\end{equation}

To determine $f_\pi$ we use Noether's theorem to construct the axial current
$\JA_\mu$.
Let us consider an infinitesimal axial SU(2)
transformation. It acts only on $\Sigma^1$ and $\Sigma^{K+1}$
at the ends of the moose:
\begin{equation}
\Sigma^1 \to U\Sigma^1, \quad\mbox{and}\quad \Sigma^{K+1} \to \Sigma^{K+1}
U,
\quad\mbox{where}\quad
U=\exp(i\alpha^a\tau^a/2).
\end{equation}
If the parameter $\alpha$ depends on coordinates, the Lagrangian changes
by $\delta {\cal L} = \JA^{a}_\mu \partial_\mu \alpha^a$. On
the other hand from (\ref{L}) one finds
\begin{equation}
\delta{\cal L}_{\pi^2} =
f\Tr(\partial_\mu\Pi)\tau^a(\partial_\mu\alpha^a) 
= f\partial_\mu\pi^a(\partial_\mu\alpha^a),
\end{equation}
which means $\JA_\mu=f\d_\mu\pi^a$, i.e., $f_\pi=f$.  Equation~(\ref{f}) 
becomes
\begin{equation}
\frac4{f_\pi^2} = \sum_{k=1}^{K+1}\frac1{f_k^2}\,.
\lb{fpi}
\end{equation}

It is a simple exercise to verify that for 
$K=0,1,2$ this general formula is in
agreement with corresponding results in these theories. 
It is
also perhaps useful to observe that sending $f_k\to\infty$ on one of
the links effectively sets gauge fields on the ends of this
link equal to each other ($(gA)^k=(gA)^{k+1}$), effectively
eliminating this link and reducing $K$ by one. The formula
\eq{fpi} obviously reflects this reduction --- the corresponding
term $1/f_k^2$ drops out.


\subsection{Vector meson mass spectrum $m_n$}

In our gauge, the vacuum is $\Sigma^k=1$ for all $k$.
Expanding to second order in $A^k_\mu$, we find the terms that determine
the masses of the vector mesons:
\begin{equation}\lb{Lmass}
{\cal L}_{A^2} = \sum_{k=1}^{K+1} 
f_k^2\Tr[ (gA_\mu)^{k-1} - (gA_\mu)^{k} ]^2 
\equiv
\sum_{\substack{k=1\\k'=1}}^{K}   (M^2)_{kk'} \Tr A^k_\mu A^{k'}_\mu\,.
\end{equation}
The mass matrix can be diagonalized by using an orthogonal matrix
$b_n^k$ satisfying
\begin{equation}\lb{mbb}
\sum_{\substack{k=1\\k'=1}}^{K} (M^2)_{kk'} b^k_m b^{k'}_n 
= m_n^2 \delta_{mn};
\qquad b^T b = b b^T = 1.
\end{equation}
In terms of new vector fields $\alpha_\mu^n$ defined as
\begin{equation}\lb{Abalpha}
A_\mu^k = \sum_n b_n^k \alpha_\mu^n
\qquad \mbox{or}\quad
\alpha_\mu^n = \sum_k  b_n^k A_\mu^k,
\end{equation}
the mass term (\ref{Lmass}) is diagonal. 

Using Eq.~\eq{Lmass} and 
the orthogonality of $b^k_n$,
we can write the equation determining $b_n^k$ and $m_n$ as
\begin{equation}\lb{evdisc}
f_{k+1}^2[(gb_n)^{k+1}-(gb_n)^k] - f_k^2[(gb_n)^k-(gb_n)^{k-1}]
= - \frac{m_n^2 b_n^k}{g_k}\,.
\end{equation}
This is essentially a discretized version of a Sturm-Liouville
problem.  We shall write the corresponding differential equation in section
\ref{sec:continuum} when we consider the continuum limit $K\to\infty$.
We shall also use the discrete equation \eq{evdisc} in section
\ref{sec:VV}.

Without solving Eq.~(\ref{evdisc}), we can conclude right away that 
there is a tower
of eigenvalues $m_n$, $n=1,2,\ldots, K$, corresponding to the masses
of vector and axial vector mesons.  
The lowest $n=1$ and $n=2$ states correspond to the $\rho$ and $a_1$ mesons.
Moreover, states with opposite 
parity alternate in the spectrum:
\begin{equation}\lb{parity}
\begin{split}
  n&=1,3,\ldots \quad (\rho, \rho', \ldots):\qquad b_n^k = +b_n^{K+1-k},\\
  n&=2,4,\ldots \quad (a_1, a_1', \ldots):\qquad b_n^k = -b_n^{K+1-k},
\end{split}
\end{equation}

Odd $n$ states correspond to vector mesons and even $n$ to axial vector
mesons.  In the real world, the trend of alternating parity can be seen in the
hadronic spectrum for the few first vector and axial vector states.

\subsection{Vector meson-pion-pion coupling $g_{n\pi\pi}$}

Let us compute the coupling of $n$th vector meson to a pion pair, 
$g_{n\pi\pi}$.
Expanding the Lagrangian in $A$ and $\pi$, isolating $A\pi\pi$
terms, and using Eq.~(\ref{Abalpha}), we find
\begin{equation}
\begin{split}
{\cal L}_{A\pi\pi} =\, & 
i \sum_{k=1}^{K+1} \frac{f_\pi^2}{ 4f_k^2}
\Tr [\partial_\mu \Pi,\Pi ] [ (gA_\mu)^{k-1} + (gA_\mu)^k ]\\
 &= 
i \sum_{n=1}^K \sum_{k=1}^{K+1} \frac{f_\pi^2}{ 4f_k^2}
[ (gb_n)^{k-1} + (gb_n)^k ]\Tr [\partial_\mu \Pi,\Pi ]\alpha^n_\mu \,.
\end{split}
\end{equation}
Recall that we use the matrix notation where
$\alpha=\alpha^a\tau^a/2$ and $\Pi=\pi^2\tau^a/2$.  If
we normalize $g_{n\pi\pi}$ so that the relevant coupling is
\begin{equation}
{\cal L}_{\alpha\pi\pi} = -g_{\alpha\pi\pi} \epsilon^{abc} 
\partial_\mu \pi^a \pi^b \alpha^c_\mu\,,
\qquad \alpha=\rho,\rho',\ldots
\end{equation}
then
\begin{equation}\lb{gnpipiK}
g_{n\pi\pi} = \sum_{k=1}^{K+1} \frac{f_\pi^2}{ 4f_k^2}
\frac12[  (gb_n)^{k-1} + (gb_n)^k ].
\end{equation}
Note that this $n\pi\pi$ coupling vanishes for axial mesons
$n=2,4,\ldots$,  as
required by parity, because their ``wave functions'' $b^n_k$
are odd under $k\to K+1-k$ (Eq.~\eq{parity}).

\subsection{Vector meson decay constants $g_{nV}$ and $g_{nA}$}
\lb{sec:VV}

We define the decay constants for the vector and axial vector mesons via
the matrix elements of the vector and axial vector currents between the
vacuum and the one-meson states,
\begin{subequations}
\begin{eqnarray}
  \<0|\JV^a_\mu(0)|\alpha_n^b(p,\epsilon)\> 
   &=& g_{nV} \delta^{ab}\epsilon_\mu\,,\\
  \<0|\JA^a_\mu(0)|\alpha_n^b(p,\epsilon)\> 
  &=& g_{nA}\delta^{ab}\epsilon_\mu\,.
\end{eqnarray}
\end{subequations}
Here $|\alpha_n^b(p,\epsilon)\>$ is a single-particle state of
the $n$-th vector boson ($\alpha_1=\rho$, $\alpha_2=a_1$, etc.)
with isospin $b$ and polarization $\epsilon$.  Both $g_{nV}$ and $g_{nA}$
have the dimension of [mass$^2$].
$g_{nV}=0$ for axial vector mesons ($n$ even) and
$g_{nA}=0$ for vector mesons ($n$ odd).  

It is convenient to compute $g_{nV}$ by looking at the
vector current-current correlator $\langle \JV_\mu(x)
\JV_\nu(0)\rangle$.  The residues at the poles are easily related to
$g_{nV}$.  The correlator can be obtained by gauging
the corresponding SU(2)$_V$ transformation and differentiating the action 
with respect to the
gauge field $B_\mu$:
\begin{equation}
\langle  \JV^a_\mu(x)\JV^b_\nu(0)\rangle
\equiv
i\langle0| T(\, \JV^a_\mu(x)\JV^b_\nu(0) \,)|0\rangle
=
-\frac{\delta^2 W_{\rm vac}[B_\mu]}{\delta B^a_\mu(x)\delta B^b_\nu(y)}
\,,
\end{equation}
were $W_{\rm vac}[B_\mu]$ is the vacuum energy functional in the
presence of the external field $B$.
The SU(2)$_V$ transformation only affects the two $\Sigma$ links
at the ends of the chain according to Eq.~\eq{usigmau}:
\begin{equation}
\Sigma^1 \to U\Sigma^1 \quad\mbox{and}\quad \Sigma^{K+1} \to \Sigma^{K+1}
U^\dag,
\end{equation}
and therefore the terms containing the gauge field $B_\mu$ are only from
$k=1$ and $k=K+1$:
\begin{equation}\label{LB}
\begin{split}
{\cal L}_B =\,& f_1^2 \Tr |\partial_\mu\Sigma^1 - iB_\mu\Sigma^1 
+ i\Sigma^1(gA_\mu)^1 |^2\\
  &
+ f_{K+1}^2 \Tr |\partial_\mu\Sigma^{K+1} - i(gA_\mu)^K \Sigma^{K+1}
+ i\Sigma^{K+1} B_\mu|^2.
\end{split}
\end{equation}
Keeping only terms bilinear in the fields we find (the term $\Pi B$ is
absent due to parity):
\begin{equation}
{\cal L}_{A^2,AB,B^2} = f_1^2 ( B^a_\mu )^2 
- B^a_\mu f_1^2[ (gA^a_\mu)^1 + (gA^a_\mu)^K ]
+ {\cal L}_{A^2} + {\cal L}_{F^2}\,.
\end{equation}
Extremizing the action with respect to $A$ at fixed $B$ and then taking
the second derivative
with respect to $B$ we find (this is also equivalent to
taking the Gaussian integral over $A$ and then differentiating
the logarithm of this integral)
\begin{equation}
\langle \JV^a_\mu(x) \JV^b_\nu(y) \rangle
=
 2 f_1^2 \eta_{\mu\nu}\delta_{xy}\delta^{ab}
- f_1^4\langle [ (gA^a_\mu)^1 + (gA^a_\mu)^K ](x) 
[ (gA^b_\nu)^1 + (gA^b_\nu)^K ](y) \rangle,
\end{equation}
where the $\langle A A \rangle$ is the propagator of $A$,
i.e., the inverse of the quadratic form found in  
${\cal L}_{A^2} + {\cal L}_{F^2}$. Diagonalizing this expression using
Eq.~(\ref{Abalpha}) and performing a Fourier transformation with respect to the
four-dimensional coordinate $x$, we find
\begin{equation}\lb{VV}
\langle \JV^a_\mu(q) \JV^b_\nu(-q) \rangle
= 
2f_1^2 \eta_{\mu\nu}\delta^{ab}
-
\sum_{n=1}^K \frac{g_{nV}^2}
{-q^2+m_n^2}\delta^{ab}\left(\eta_{\mu\nu} - \frac{q_\mu q_\nu}{m_n^2}\right),
\end{equation}
where the decay constants $g_{nV}$ are determined to be
\begin{equation}\lb{gnv}
g_{nV} = f_1^2 [ (gb_n)^1 + (gb_n)^K ]. 
\end{equation}
Note that $g_{nV}=0$ for axial mesons for which $b_n^1=-b_n^K$.

One can also use Eq.~(\ref{evdisc}) together with Eq.~(\eq{gnv})
to write a different representation for $g_{nV}$:
\begin{equation}\lb{gnvK}
g_{nV} = f_1^2 [ (gb_n)^1 + (gb_n)^K ] = m_n^2\sum_{k=1}^K \frac{b_n^k}{g_k}
\,.
\end{equation}
It is perhaps easier to look at this equation as a discretized
version of integration by parts as we shall do in section~\ref{sec:continuum}.%
\footnote{Equation (\ref{gnvbg}) or (\ref{gnvK}) can be 
also understood in the following way (the reader might
recognize the discussion given in Refs.~\cite{Kroll,Sakurai}). One should
realize that because of the mixing of $B$ with $A$ the actual photon
is not the field $B$, but a linear combination of $B$ and all $A$
that leaves vacuum $\Sigma^k=1$ invariant. This is similar to
the mixing of the standard model hypercharge boson and weak isospin
vector boson to produce the photon. The corresponding linear
combination of the $A$ fields has a ``wave function'' proportional
to $1/g_k$ (each $A_k$ enters with weight $1/g_k$). The mixing of the
actual photon is now entirely through derivative terms in the
expansion of ${\cal L}_{F^2}$. The corresponding coefficients are
given by the overlap of the photon ``wave function'' $1/g$ and
the $n$th vector meson ``wave function'' $b_n$. The factor $m_n^2$
is from the derivatives evaluated using the equation of motion for the
$n$th meson.}

The calculation of the decay constants of the axial vector mesons $g_{nA}$
is completely analogous.  We introduce an auxiliary gauge
field $\tilde B$ coupled to the axial current $\JA$:
\begin{equation}
\begin{split}
{\cal L}_{\tilde B} &= 
f_1^2 \Tr |\partial_\mu\Sigma^1 - i\tilde B_\mu\Sigma^1 
+ i\Sigma^1(gA_\mu)^1 |^2 \\
&\quad+ 
 f_{K+1}^2 \Tr |\partial_\mu\Sigma^{K+1} - i(gA_\mu)^K \Sigma^{K+1}
- i\Sigma^{K+1} \tilde B_\mu|^2.
\end{split}
\end{equation}
In addition to the $A\tilde B$ mixing and $\tilde B^2$ contact terms
there is now also the mixing with pion field $\Pi\tilde B$.
Differentiating the logarithm of the partition function
twice, we obtain
\begin{equation}\lb{AA}
\langle \JA^a_\mu(q) \JA^b_\nu(-q) \rangle
= 
 2f_1^2 \eta_{\mu\nu}\delta^{ab}
-\sum_{n=1}^K \frac{g_{nA}^2}
{-q^2+m_n^2}\delta^{ab}\left(\eta_{\mu\nu} - \frac{q_\mu q_\nu}{m_n^2}\right)
- f_\pi^2 \frac{q_\mu q_\nu}{q^2}\delta^{ab},
\end{equation}
where
\begin{equation}\lb{gna}
g_{nA} = f_1^2 [ (gb_n)^1 - (gb_n)^K ].
\end{equation}
Note that, as expected, $g_{nA}=0$ for vector mesons for which $b_n^1=b_n^K$.

\subsection{Spectral sum rules}
\label{sec:Weinberg}

Weinberg \cite{Weinberg-sum}
has derived two sum rules for the weighted integrals 
of the difference of spectral functions of the vector 
and axial vector current correlators,
$\langle\JV\JV\rangle$ and $\langle\JA\JA\rangle$.
We shall now verify that both sum rules hold
by using transversality
of the current-current correlators.

From Eq.~(\ref{gnvK}) it is easy to show that the $\<\JV\JV\>$
correlator is transverse. Transversality requires that the contact
term in Eq.~(\ref{VV}) is related to the pole terms through the following
sum rule:
\begin{equation}\lb{gnvsum}
2f_1^2 = \sum_{n=1}^K \frac{g_{nV}^2}{m_n^2}\,.
\end{equation}
This can be checked by using Eq.~(\ref{gnvK}) and orthogonality of 
$b_n^k$. We indeed find
\begin{equation}
\sum_{n=1}^K \frac{g_{nV}^2}{m_n^2}= \sum_{n=1}^K \sum_{k=1}^K 
f_1^2 [ (gb_n)^1 + (gb_n)^K ]\frac{b_n^k}{g_k}= 2 f_1^2.
\end{equation}
By using Eq.~(\ref{gnvsum}) one can rewrite the correlator in a manifestly
transverse form,
\begin{equation}\lb{VVPi}
\begin{split}
& \langle \JV^a_\mu(q) \JV^b_\nu(-q) \rangle
= \Pi_V(-q^2)\delta^{ab} (-q^2\eta_{\mu\nu}+ q_\mu q_\nu); \\
& \quad\mbox{where}\qquad
\Pi_V(Q^2)=\sum_{n=1}^K \frac{g_{nV}^2}
{m_n^2(Q^2+m_n^2)}\,,\qquad Q^2=-q^2.
\end{split}
\end{equation}

The transversality of the $\<\JA\JA\>$ correlator~(\ref{AA}) amounts to the 
following sum rule:
\begin{equation}\lb{gnasum}
2f_1^2 - f_\pi^2 = \sum_{n=1}^K \frac{g_{nA}^2}{m_n^2}\,.
\end{equation}
By comparing Eqs.~\eq{gnvsum} and \eq{gnasum} we conclude that
\begin{equation}\lb{weinberg1}
\sum_{n=1}^K 
\left(\frac{g_{nV}^2}{m_n^2} - \frac{g_{nA}^2}{m_n^2}\right) = f_\pi^2\,,
\end{equation}
which is one of Weinberg's sum rules.%
\footnote{Weinberg's sum rules \cite{Weinberg-sum}
involve the spectral functions
$$\rho_V(\mu^2) \equiv \frac{\mu^2}\pi {\rm Im } \Pi_V(-\mu^2-i0)$$
where
$\Pi_V$ is defined via $\langle\JV\JV\rangle$
in Eq.~\eq{VVPi}. A similar equation defines $\rho_A$.
The sum rules state that
$$(i) \int [\rho_V(\mu^2) - \rho_A(\mu^2)]\mu^{-2}d\mu^2=f_\pi^2;
\qquad
(ii) \int [\rho_V(\mu^2) - \rho_A(\mu^2)]d\mu^2 = 0.
$$
In our theory, according to Eq.~\eq{VVPi},
 $\rho_{V,A}= \sum_n g_{nV,A}^2 \delta(-\mu^2+m_n^2)$.
}
This sum rule holds for any $K$. Note that, for $K\to\infty$ 
both sums \eq{gnvsum} and \eq{gnasum} must diverge, since
$f_k$ must become infinite at the ends of the moose (to ensure
convergence in Eq.~\eq{fpi}). However,
their difference is finite.

The second Weinberg sum rule
\begin{equation}\label{weinberg2}
\sum_{n=1}^K (g_{nV}^2 - g_{nA}^2)=0
\end{equation}
also holds. It is easy to prove by using the definitions \eq{gnv} 
and \eq{gna} and the orthogonality of $b_n^k$:
\begin{equation}
\sum_{n=1}^K (g_{nV}^2 - g_{nA}^2)= 4 f_1^2
\sum_{n=1}^K (gb_n)^1 (gb_n)^K = (2 f_1 g_1)^2 \delta_{1K},
\end{equation}
which vanishes for all $K>1$. In the case $K=1$, there
is only one meson -- $\rho$, and no axial mesons at all.

\subsection{Pion form factor and VMD}

The pion form factor (defined to be the isovector part of the electromagnetic 
form factor),
\begin{equation}
\langle \pi^a(p') | \JV^c_\mu(0) |\pi^b(p)\rangle = 
G_{V\pi\pi}(q) \epsilon^{abc} (p+p')_\mu\,
\end{equation}
can be found isolating terms linear in $B$ in the Lagrangian~(\ref{LB}). 
There are
two contributions to the form factor -- the direct interaction,
given by the term $B\pi\partial\pi$ in the Lagrangian and
the interaction mediated by vector mesons given by the
$AB$ mixing terms and the couplings $A\pi\partial\pi$.
One finds
\begin{equation}\lb{gvpipi}
G_{V\pi\pi}(q) = \frac{f_\pi^2}{4f_1^2} +
\sum_n \frac{g_{nV}g_{n\pi\pi}}{Q^2+m_n^2}\,.
\end{equation}
Using the expressions~(\ref{gnvK}) and
(\ref{gnpipiK}) for $g_{nV}$ and $g_{n\pi\pi}$
and the orthogonality of the matrix $b_n^k$, the
sum rule related to the total charge of pion can be verified,
\begin{equation}\lb{gvpipi0}
G_{V\pi\pi}(0)=\frac{f_\pi^2}{4f_1^2} +
\sum_n \frac{g_{nV}g_{n\pi\pi}}{m_n^2} = 1.
\end{equation}
If we understand VMD as the statement that $G_{V\pi\pi}(q)$ is saturated
by a sum over resonances (i.e., dominance by the whole tower of mesons), 
then in our model
VMD is valid when the contribution of the
direct interaction is negligible, ${f_\pi^2}/{4f_1^2}\ll1$.
Thus VMD is a natural consequence of the $K\to\infty$ limit
(due to Eq.~\eq{fpi}).  A stronger statement that $G_{V\pi\pi}(q)$ is saturated
by a single $\rho$ pole is not, in general, valid (see, however, 
section~\ref{sec:cosh}).

\section{$K\to\infty$ and continuum limit}
\lb{sec:continuum}

In the preceding section we derived formulas that are valid
for an arbitrary $K$. Now we wish to consider the limit $K\to\infty$.
In this limit the expressions that we found 
can be simplified, provided that $f_k$ and
$g_k$ are sufficiently smooth functions of $k$. In this case
we can consider replacing the discrete variable $k$ by a continuum
variable that we shall call $u$:
\begin{equation}
u = \left( k-\frac K2 \right)a\,,
\end{equation}
Here $a$ plays the role of the ``lattice spacing.''
If the limit $K\to\infty$ is performed in the following way,
\begin{equation}
K\to\infty \quad{\rm and}\quad  a\to0, \qquad 
Ka\equiv2u_0 \quad {\rm fixed},
\end{equation}
then $u$ becomes a continuum replacement for $k$.
If $f_k$ and $g_k$ are smooth functions of $k$,
we can also replace them by functions of $u$,
\begin{subequations}
\begin{eqnarray}\lb{fkfu}
  af_k^2  &=& 
   f^2\left(\Bigl( k-\frac K2 \Bigr)a\right)={f^2(u)}\,,\\
  ag_k^2  &=& 
   g^2\left(\Bigl( k-\frac K2 \Bigr)a\right)={g^2(u)}\,.
\end{eqnarray}
\end{subequations}
For the resonance ``wave functions'' $b_n^k$ that vary smoothly we can
write
\begin{equation}
\frac1{\sqrt a} b_n^k = b(u),
\end{equation}
so that orthogonality of the matrix $b_n^k$ translates into orthonormality
of the functions $b_n(u)$,
\begin{equation}
\int_{-u_0}^{u_0}\! du\, b_n(u) b_{n'}(u) = \delta_{nn'}\,.
\end{equation}
The wave functions of
sufficiently high resonances with $n\sim K$ cannot be expected to be
smooth, so they must be treated discretely. We shall always be
interested in a finite number of lowest resonances, while $K\to\infty$.

\subsection{Physical observables}
\lb{sec:contform}

Let us now write continuum limits for the main formulas we have
derived in the preceding section. From Eq.~\eq{fpi},
\begin{equation}\lb{fpiu}
\frac 4{f_\pi^2} = \int_{-u_0}^{+u_0}\! \frac {du}{f^2(u)}\,.
\end{equation}

From Eq.~\eq{evdisc},
\begin{equation}\lb{eigen}
g(f^2(gb_n)')' = - m_n^2 b_n.
\end{equation}
with Dirichlet boundary conditions $b_n(\pm u_0)=0$ (since we
 set $A_\mu^0=A_\mu^{K+1}=0$).

From Eq.~\eq{gnpipiK},
\begin{equation}\lb{gnpipiu}
g_{n\pi\pi} = \frac{f_\pi^2}4
  \int_{-u_0}^{+u_0}\!\frac{du}{f^2(u)}\,g(u)b_n(u).
\end{equation}

From Eq.~\eq{gnv}, using the fact that $(gb)^0=(gb_n)^{K+1}=0$,
\begin{equation}
g_{nV} = -[f^2(u) (g(u)b(u))']_{-u_0}^{+u_0}.
\end{equation}
%
By using Eq.~\eq{eigen}, we find the continuum limit
of Eq.~\eq{gnvK}:
\begin{equation}\lb{gnvbg}
g_{nV} = -[f^2(u) (g(u)b(u))']_{-u_0}^{+u_0}= m_n^2\int_{-u_0}^{+u_0} 
  \!du\, \frac{b_n(u)}{g(u)}\,.
\end{equation}
Analogously, Eq.~(\ref{gna}) becomes
\begin{equation}\label{gnabg}
  g_{nA} = [f^2(u) (g(u)b(u))']|_{+u_0}+[f^2(u) (g(u)b(u))']|_{-u_0}\,.
\end{equation}
It is very interesting that the physical observables 
we calculated are all well behaved 
in the continuum limit $K\to\infty$, $a\to0$ 
(provided the corresponding integrals over $u$ converge).
For reference, the equations for other vertex couplings are presented in
 Appendix \ref{app:mnp}.

\subsection{$d=4+1$ and dimensional deconstruction}
\label{sec:5dgauge}

Our long-moose theory with $K\gg1$ can be also considered
as a discretized (or deconstructed) five-dimensional continuum
gauge theory in curved spacetime.%
\footnote{Deconstruction of gauge theories in curved space was considered
in Refs.~\cite{Sfetsos,FalkowskiKim,WeinerRandall}.}
The variable $u$ plays the role of the fifth, deconstructed, 
dimension.
The smoothly varying fields $\Sigma$'s can be interpreted as 
the link variables along the fifth dimension $u\equiv x^5$,
\begin{equation}\lb{SigmaA5}
  \Sigma^k \approx 1 + iaA_5(u),
\end{equation}
For this equation and for remainder of section \ref{sec:continuum} 
we shall make a temporary
switch of notations, absorbing the gauge coupling constants $g$ into
the fields $A$: $gA\to A$.
Then the action~\eq{L}
can be written in the 5d notations as
\begin{equation}\label{SfgF}
  S = -\Tr\int\!du\,d^4x\,\Bigl(-f^2(u)F_{5\mu}^2
      +\frac1{2g^2(u)} F_{\mu\nu}^2\Bigr).
\end{equation}

We now compare this action to the action of a gauge field in a
background of curved spacetime and a dilaton field. In the following,
$\dg$ denotes the determinant of the metric tensor. The action is
taken in the form:
\begin{equation}\lb{5dS}
  S = -\frac1{2g_0^2}\Tr\int\!d^5x\, \sqrt{\dg}\, e^{-2\phi} 
  F_{\hat{\mu}\hat{\nu}} F^{\hat{\mu}\hat{\nu}},
\end{equation}
where $\hat{\mu},\hat{\nu}$ are 5d Lorentz indices.  The coupling to
the dilaton field is written so that the effective gauge coupling is
$g=g_0e^\phi$.  In our simple model we consider the metric and the
dilaton as classical background fields with no dynamics of their
own.  Taking the dilaton field to be dependent only on the fifth
coordinate $u$, $\phi=\phi(u)$, and the metric to be of the warped
form,
\begin{equation}\label{warped}
  ds^2 = -du^2 + e^{2w(u)}\eta_{\mu\nu} dx^\mu dx^\nu,
\end{equation}
the action (\ref{5dS}) can be expanded as
\begin{equation}\lb{5dSexp}
  S = -\frac1{2g_0^2}\Tr\int\!d^5x\, \Bigl(-2e^{2w-2\phi}F_{5\mu}^2 
      + e^{-2\phi}F_{\mu\nu}^2\Bigr).
\end{equation}
Equation~(\ref{5dSexp}) coincides with Eq.~(\ref{SfgF}) if one makes the
following identification:
\begin{subequations}\lb{fgwarp}
\begin{eqnarray}
  f^2(u) &=& \frac1{g_0^2} e^{2w-2\phi},\\
  g^2(u) &=& g_0^2 e^{2\phi}.
\end{eqnarray}
\end{subequations}
Notice that the warp factor $e^{2w}$ is equal to $f^2(u)g^2(u)$,
i.e.,
\begin{equation}\lb{metric}
ds^2 = -du^2 + f^2g^2\eta_{\mu\nu} dx^\mu dx^\nu.
\end{equation}
It is also easy to see that the wave equation for the spin-1
mesons \eq{eigen} is one of the Yang-Mills/Maxwell equations
for the 5d (massive) gauge field in the curved background.
Notice also that the pion field~(\ref{Sigmaproduct}) is now the
Wilson line stretching between the two boundaries,
\begin{equation}\label{Sigmacont}
  \Sigma(x) = P\exp\Bigl(i\int\!du\, A_5(u,x)\Bigr)\,.
\end{equation}

\subsection{AdS/CFT connection}

We now show that correlators of conserved currents in our
theory can be computed by using a prescription essentially identical
to the AdS/CFT one.  Namely, the generating functional for the
correlation functions of the currents is equal to the action of a
solution to the classical field equations, with the sources serving as
the boundary values for the classical fields.

Recall our calculation of the current-current correlators in Section
\ref{sec:VV}. Instead of the vector field $B_\mu$ and $\tilde B_\mu$
let us introduce two
separate fields $A^L_\mu$ and $A^R_\mu$, corresponding to gauging
the SU(2)$_L$ and SU(2)$_R$ global symmetries of the theory.
The appearance of these fields modify the first and last terms in the moose,
\begin{equation}
  {\cal L} = f_1^2\Tr|\d_\mu\Sigma^1-iA^L_\mu\Sigma^1+i\Sigma^1 A^1_\mu|^2
      + \cdots +
      f_{K+1}^2\Tr|\d_\mu\Sigma^{K+1} -iA^K_\mu\Sigma^{K+1}
      + i\Sigma^{K+1} A^R_\mu|^2
\end{equation}
In this section we also absorb the coupling $g_k$ into the field $A$.
Remember that there are no dynamical fields associated with the ends 
of the moose $k=0$ and $K+1$. 
We can treat the ends of the moose more equally with the other points
by thinking that the values of the field $A^k$ at the ends of the moose,
at $k=0$ and $K+1$, are fixed at given values:
\begin{equation}
A_\mu^0 = A^L_\mu  \quad \mbox{and} \quad A_\mu^{K+1}=A^R_\mu.
\end{equation}
If the field $A^k$ is smooth, we can translate this
into the continuum limit by setting boundary conditions on the
continuous 5d field $A_\mu(u)$:
\begin{equation}\lb{bc}
A_\mu(- u_0) = A_\mu^L 
\quad\mbox{and}\qquad  
A_\mu(+ u_0) = A_\mu^R.
\end{equation}

At tree level, the generating functional is thus equal to
\begin{equation}\label{AdSCFT}
  Z[A^L_\mu,\, A^R_\mu] = e^{iS_{\textrm{cl}}[A_\mu^{\textrm{cl}}]}
\end{equation}
where $A_\mu^{\textrm{cl}}$ is the solution to the classical field
equation that satisfies the boundary conditions (\ref{bc}).  This
formula is of the same form as the formula for AdS/CFT correspondence:
the sources for the boundary theory (in our case $A^{L,R}_\mu$) serve
as the boundary values for the bulk field.  In particular, in order to
compute the correlation functions for the conserved currents 
$\JL_\mu = \frac12(\JV_\mu + \JA_\mu)$ or 
$\JR_\mu = \frac12(\JV_\mu - \JA_\mu)$
one just needs to differentiate the classical action with
respect to the corresponding boundary values, e.g.,
\begin{equation}
  \<\JL_\mu(x) \JL_\nu(y)\> = 
\frac{\delta^2 S_{\textrm{cl}}[A_\mu^{\textrm{cl}}]}
  {\delta A^L_\mu(x) \delta A^L_\nu(y)}
\end{equation}

The fact that we have arrived at an AdS/CFT-like formula (\ref{AdSCFT})
makes one wonder if the hidden local symmetry models for the $\rho$ and
$a_1$ vector mesons \cite{HLS,HLSa1,HLSreview} 
are (very coarsely) discretized versions of a 5d
theory dual to QCD.  This could explain why these models enjoy certain
phenomenological success, and why in the $K=2$ model \cite{HLSa1}
one is driven to choose the parameters so that it becomes a moose
theory (i.e., nearest neighbor local).

A difference from the usual AdS/CFT correspondence is that there are
{\em two} boundaries in the open moose theory.  However, if so
desired, one can reformulate the 5d theory~(\ref{5dS}) and
(\ref{warped}) in the spatial region $0<u<u_0$ (which is one half of
the original $-u_0<u<u_0$) at the price of having two gauge fields
obeying a matching condition at $u=0$.  Then the spacetime will have 
only one boundary at $u=u_0$.

\section{Exactly solvable examples and phenomenology}
\label{sec:examples}

So far, our discussion has been general and valid for any choice of 
$f_k$ and $g_k$.
In this section we shall consider two concrete realizations of the
open-moose theory.  Our goal is to illustrate the general formulas,
and to compare the results with the phenomenology of vector mesons.
The two examples are chosen because they are exactly solvable:
the spectrum of the vector mesons and the
coupling constants can be found in the closed form.
The first example is also the simplest possible model, 
but it has a significant physical
drawback that we point out at the end. We think nevertheless that
it is a useful reference point for comparison and for understanding
the robustness/sensitivity of the results towards the change
of the background parameters $f(u)$ and $g(u)$.

\subsection{Example I: flat background}

Consider a moose with parameters $f_k$ and $g_k$ independent
of $k$.\footnote{ Such a theory can be easily solved even for finite $K$,
but we shall only consider $K\gg1$.}
In the continuum limit $K\gg1$ the corresponding functions are
therefore constant,\footnote{The choice of $u_0$ does not affect
the results; it is equivalent to rescaling $f$ and $g$.}
\begin{equation}
f(u)=f,\qquad g(u)=g, \quad |u|<u_0=1.
\end{equation}
Let us now apply general formulas from section \ref{sec:contform}
to determine the properties of this theory in terms of
the parameters $f$ and $g$.
From \eq{fpiu}
\begin{equation}
f_\pi^2 =  2 f^2.
\end{equation}
%
The spectrum and wave functions of the spin-1 mesons are given by
Eq.~\eq{eigen}, which becomes
%
\begin{equation}
b_n'' + \frac{m_n^2}{f^2g^2}b_n = 0, \quad b_n(\pm1)=0.
\end{equation}
This means
\begin{equation}\lb{mn1}
b_n(u) = \sin\left(\frac{\pi n}2 (u+1)\right);
\qquad 
m_n = \frac{\pi fg}2 n;
\qquad
n=1,2,\ldots.
\end{equation}
The few first wave functions are plotted in Fig.~\ref{fig:sin}.
\begin{figure}
\begin{center}
\epsfig{file=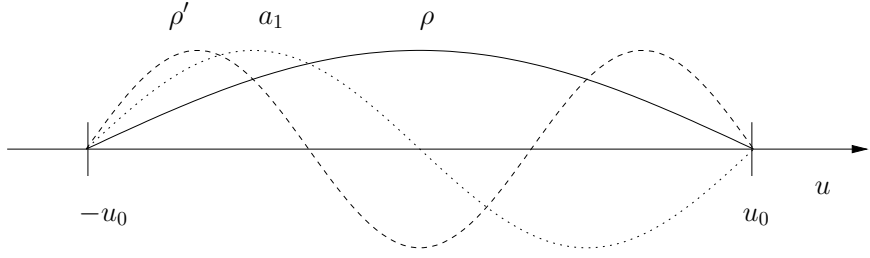,width=.7\textwidth}
\end{center}
\caption[]{A few first wave functions in the flat background}
\lb{fig:sin}
\end{figure}

From Eq.~\eq{gnpipiu}, for $n=1,3,\ldots$,
\begin{equation}
g_{n\pi\pi}
=
\frac {f_\pi^2}{\pi} \frac {g}{f^2} \frac1n
=
\frac{2}\pi g \frac1n;
\qquad 
n=1,3,\ldots .
\end{equation}
Consider the $\rho$ meson, $n=1$.
The ratio of $m_\rho^2$ to $g_{\rho\pi\pi}^2f_\pi^2$ is dimensionless
and is equal to
\begin{equation}\label{KSRF}
\frac {m_1^2} {g_{1\pi\pi}^2f_\pi^2} = \frac{\pi^4}{32} \approx 3.04.
\end{equation}
The coupling $g_{\rho\pi\pi}$ can be found from the width of the $\rho$,
which decays predominantly to two pions:
$\Gamma_\rho=g_{\rho\pi\pi}^2m_\rho v_\pi^3/(48\pi)$, where $v_\pi$ is
the velocity of the final-state pions.  Using $\Gamma_\rho\approx150$
MeV, we find the ratio~(\ref{KSRF}) to be around 1.9 in Nature.  
For comparison,
the KSRF relation~\cite{KS,RF} corresponds to this ratio being
equal to 2, and the value in Georgi's vector limit (i.e., $K=1$ moose
theory) is 4~\cite{Georgi}.  
Therefore, our model would underpredict $\Gamma_\rho$ from
experimental $m_\rho$ and $f_\pi$.

The decay constants $g_{nV}$ and $g_{nA}$ are given by Eqs.~\eq{gnvbg} and
(\ref{gnabg})
and are equal to
\begin{equation}\lb{gnv1}
g_{nV,A} = m_n^2 \frac{4}\pi \frac1g \frac 1n
=
\pi  f^2  g n.
\end{equation}
In Eq.~(\ref{gnv1}) $g_{nV,A}$ refers to $g_{nV}$ for odd $n$'s and 
$g_{nA}$ for even $n$'s.
For $n=1$, we find $g_{\rho V}=\sqrt2f_\pi m_\rho$.  We can now predict
the rate of the electromagnetic decay $\rho^0\to e^+e^-$, using
$\Gamma(\rho^0\to e^+e^-)=\frac43\pi\alpha^2g_{\rho V}^2m_\rho^{-3}$
and the experimental values for $f_\pi$ and $m_\rho$.  We find
$\Gamma(\rho^0\to e^+e^-)\approx 5.0$ keV, which is somewhat smaller than
the measured value $6.85\pm0.11$~keV~\cite{PDG}.

It is also interesting to consider the contribution of
the $\rho$ meson to pion form factor at $q=0$ (Eq.~\eq{gvpipi0}),
\begin{equation}
\frac { g_{1V} g_{1\pi\pi} }{ m_1^2 } = \frac8{\pi^2} \approx 0.81.
\end{equation}
Thus the single $\rho$-meson dominance holds to within 20\%.
The VMD is, however, exact if all vector mesons are included 
in the limit $K\to\infty$.
Indeed the direct pion-photon interaction in Eq.~\eq{gvpipi} vanishes: 
$f_\pi^2/4 f_1^2 = af_\pi^2/4f^2 = a/2\to0$.%
 \footnote{This can be verified also 
by summing the contributions from all vector
mesons in  \eq{gvpipi0}. Each contribution is proportional to
  $1/n^2$ and $\sum_{n=1,3,\ldots} 1/n^2 =\pi^2/8$.}

The drawback of this model is that it fails to satisfy the asymptotic
condition on $\Pi_V(Q^2)$ that follows from QCD:
$\Pi_V(Q^2)\sim
N_c\log (Q^2)$ when $Q^2\to\infty$. Instead, $\Pi_V(Q^2)$
in this model vanishes as $1/\sqrt {Q^2}$ when $Q^2\to\infty$.
Indeed, according to Eq.~\eq{VVPi}, with values of $g_{nV}$ and $m_n$
found in Eqs.~\eq{gnv1} and \eq{mn1},
\begin{equation}
\Pi_V(Q^2) = \sum_{n=1,3,\ldots}
\frac{g_{nV}^2}{m_n^2(Q^2+m_n^2)} =
\frac {2f}{g Q}
\tanh\left(\frac Q{ f g}\right).
\end{equation}
We shall now consider an exactly solvable
 model that will satisfy the condition
$\Pi(Q^2)\sim\log Q^2$ at large $Q^2$.

\subsection{Example II: ``cosh'' background}
\label{sec:cosh}

This model is given by
\begin{subequations}\label{coshbg}
\begin{eqnarray}
  g(u) &=& g_5 = \textrm{const}\,,\\
  f(u) &=& \frac\Lambda{g_5} \cosh u .
\end{eqnarray}
\end{subequations}
According to Eqs.~(\ref{fgwarp}), this corresponds to a constant dilaton
background and the following background metric,
\begin{equation}\label{coshmetric}
  ds^2 = -du^2 + \Lambda^2\cosh^2 u\,\eta_{\mu\nu}dx^\mu dx^\nu.
\end{equation}
The two boundaries are located at $u=\pm\infty$.  Near the boundaries
the metric becomes asymptotically AdS$_5$.  According to the AdS/CFT
philosophy, $u$ has the physical meaning of the energy scale; large $u$'s
correspond to short distances. Therefore one can expect
that the current correlators has the conformal form at short
distance, i.e., as $Q^2\to\infty$,
\begin{equation}\label{Piexp}
  \Pi_V(Q^2), ~ \Pi_A(Q^2) \sim \log(Q^2).
\end{equation}
The main reason for choosing the background~(\ref{coshbg}) is that
$\cosh u$ is the simplest function interpolating between
$e^{-u}$ and $e^u$, and that the mass spectrum can be found exactly
(see below).  Otherwise, we have no reason to prefer this background
over any other that has two AdS$_5$ boundaries.%
\footnote{Curiously, (\ref{coshmetric}) coincides with the 
5d part of the induced metric on a probe D7 brane in
${\rm AdS}_5\times{\rm S}^5$~\cite{KruczenskiMyers}.}

Applying Eq.~(\ref{fpiu}), one finds
\begin{equation}\lb{fpiLambdag}
  f_\pi^2 = \frac{2\Lambda^2}{g_5^2}\,.
\end{equation}
The wave equation for the vector mesons is
\begin{equation}
  (\cosh^2 u\, b_n')' = -\frac{m_n^2}{\Lambda^2} b_n\,,
\end{equation}
which implies the following spectrum:
\begin{equation}\lb{mncosh}
  m_n^2 = n(n+1)\Lambda^2\, \qquad n=1,\, 2,\ldots
\end{equation}
In particular, $m_\rho^2=2\Lambda^2$ and $m_{a_1}^2=6\Lambda^2=3m_\rho^2$.  
Taking $m_\rho$ as an input, this predicts $m_{a_1}=1335$ MeV, which is not
far from the observed $1230\pm40$ MeV. 
However, the masses of higher
excitations grow faster with $n$ than in the real world.  The 5d
eigenfunctions of the vector mesons are
\begin{equation}\label{bncosh}
  b_n(u) = -c_n
  \,\frac{P^1_n(\tanh u)}
  {\cosh u}\,,\qquad
  c_n = \sqrt{\frac{2n+1}{2n(n+1)}}\,,
\end{equation}
where $P^1_n$ are the associated Legendre functions.  The first few
wave functions are (see Fig.~\ref{fig:cosh}):
\begin{subequations}
\begin{eqnarray}
b_1(u)  &=&  \frac{\sqrt3}2 \frac1{\cosh^2 u}\,,\\
b_2(u)  &=& \frac{\sqrt{15}}2\frac{\sinh u}{\cosh^3 u}\,,\\
b_3(u)  &=&  -\frac12\sqrt\frac{21}2\frac1{\cosh^2 u}\Bigl(
    \frac5{2\cosh^2 u}-2\Bigr).
\end{eqnarray}
\end{subequations}
\begin{figure}
\begin{center}
\epsfig{file=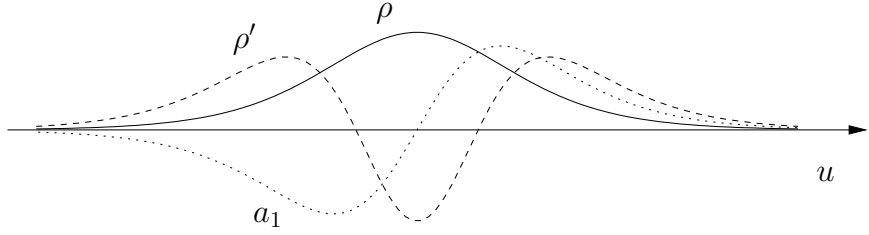,width=.7\textwidth}
\end{center}
\caption[]{A few first wave functions in the ``cosh'' background.
}
\lb{fig:cosh}
\end{figure}

In order to establish Eq.~(\ref{Piexp}), we compute
the decay constants of vector mesons from Eqs.~(\ref{gnvbg}) and 
(\ref{gnabg}),
\begin{equation}
  g_{nV,A} = \sqrt{2n(n+1)(2n+1)}\, \frac{\Lambda^{2}}{g_5}\,.
\end{equation}
The correlation function for the vector current is found from
Eq.~(\ref{VVPi}),
\begin{equation}
  \Pi_V(Q^2) = \frac{2\Lambda^2}{g_5^2}\sum_{n\textrm{ odd}} 
  \frac{2n+1}{Q^2+n(n+1)\Lambda^2}\,.
\end{equation}
At large $Q^2\gg\Lambda^2$, the sum can be replaced by an integral, which
is logarithmically divergent.%
\footnote{We perform a trivial regularization in Eq.~\eq{PiVlog},
subtracting a constant equal to $(1/g_5^2)\log (K\Lambda)^2$ for $K\gg1$. Of
course, the equation is only valid for $Q^2\ll(K\Lambda)^2$.}
One thus finds for large $Q^2$
\begin{equation}\label{PiVlog}
  \Pi_V(Q^2) = -\frac1{g_5^2}\ln(Q^2),
\qquad Q^2\gg\Lambda^2.
\end{equation}
The asymptotic behavior of $\Pi_A(Q^2)$ is the same.  Thus the 
current correlators have the correct asymptotics at large $Q^2$.  Moreover,
they obey Weinberg's sum rules, as proven in section~\ref{sec:Weinberg}.
The constraints imposed by the $Q^2\to\infty$ behavior and Weinberg's
sum rules on the masses $m_n$ and decay constants $g_{nV,A}$ are quite
nontrivial~\cite{Beane}.  It is remarkable that the open-moose 
construction generates examples that automatically satisfy these 
constraints.

One can match the asymtotics~(\ref{PiVlog}) with the result found from QCD,
\begin{equation}
  \Pi_V(Q^2) = -\frac{N_c}{24\pi^2}\ln(Q^2),
\end{equation}
where $N_c$ is the number of colors, to obtain
\begin{equation}\label{g5Nc}
  \frac1{g_5^2}
 = \frac{N_c}{24\pi^2}\,.
\end{equation}
By using this relationship between $g_5$ and $N_c$ together with
$m_\rho=\sqrt2\Lambda$, we can now
express all quantities in the model via a single mass
$m_\rho$ and the number of colors $N_c$.  A short summary is given
in Appendix~\ref{app:cosh}.
For example, for $f_\pi$ we find from Eq.~\eq{fpiLambdag}
\begin{equation}\lb{mrhofpiNc}
f_\pi^2 = \frac{N_c}{24\pi^2}m_{\rho}^2\,.
\end{equation}
For $N_c=3$
Eq.~(\ref{mrhofpiNc}) predicts $f_\pi=$ 87 MeV, rather close to the
experimental value of 93~MeV.  Interestingly, Eq.~(\ref{mrhofpiNc})
coincides with the one obtained from QCD sum rules~\cite{SVZ}.  The
large $N_c$ scaling in \eq{mrhofpiNc} also matches:
$m_\rho\sim1$, $f_\pi\sim\sqrt{N_c}$.

Another distinct feature of the model is that the pion form factor is
dominated by a single $\rho$ pole.  Indeed,
the coupling $n\pi\pi$ is found by substituting Eq.~(\ref{bncosh})
into Eq.~(\ref{gnpipiu}),
\begin{equation}
  g_{n\pi\pi} = -\frac{c_n}2 g_5 \int_{-1}^1\!d\xi\,
  \sqrt{1-\xi^2}\,P^1_n(\xi),
\end{equation}
which vanishes for all $n\ne1$. This is due to the
orthogonality of Legendre functions and the fact that
$\sqrt{1-\xi^2}=P^1_1(\xi).$  Therefore $\rho$ meson dominance is
exact for the pion form factor: $G_{V\pi\pi}(q)=(Q^2+m_\rho^2)^{-1}$.
For $n=1$ one finds
\begin{equation}
  g^2_{1\pi\pi}\equiv g^2_{\rho\pi\pi} = \frac{g_5^2}3 = \frac{8\pi^2}{N_c}
 = 
  \frac{m_\rho^2}{3f_\pi^2}\,.
\end{equation}
The KSRF ratio in this model is equal to 3. This means that the $\rho$
width is underpredicted by a factor of about 2/3.  
However, it is still interesting to compute
$\Gamma_\rho$ for arbitrary $N_c$, in the chiral limit,
\begin{equation}
  \Gamma_\rho = \frac\pi{6N_c} m_\rho\,.
\end{equation}
The rate of the electromagnetic decay $\rho^0\to e^+e^-$ in this model,
\begin{equation}
  \Gamma(\rho^0\to e^+e^-) = \frac{\alpha^2N_c}{6\pi}m_\rho\,,
\end{equation}
is equal to 6.5 keV at $N_c=3$, which is rather close to the observed
$6.85\pm0.11$~keV.
Interestingly, the prediction from QCD sum rules~\cite{SVZ} 
is close but different in this
case: $\Gamma(\rho^0\to e^+e^-)_{\rm s.r.}/\Gamma(\rho^0\to
e^+e^-)_{\cosh}= e/3$.

The phenomenology of the $a_1$ meson in this model is discussed in 
Appendix \ref{app:a1}.
The excitations with $n>2$ have an unrealistic mass spectrum in our
model, so we shall not discuss their phenomenology.

\section{Instanton $\sim$ baryon}
\label{sec:baryon}

The baryon appears in the framework of chiral Lagrangians as a
solitonic object: a Skyrmion \cite{Skyrme}.  One wonders: what is the
corresponding object in 5d that can describe the baryon.  An obvious
candidate is the instanton, which can be ``lifted'' to become a
quasiparticle in 5d.  Here we show that the instanton appears from the
point of view of 4d as a Skyrmion.  We are interested only in
topological aspects, and defer the question of stability of such a
solution to future work.\footnote{It is interesting to note
in this regard that the issue of stability of the
Skyrmion in models with a $\omega$, $\rho$, and $a_1$ mesons
has been studied \cite{AdkinsNappi,Igarashi,MeissnerZahed,ZahedBrown}.
It was determined that vector mesons not only stabilize
the Skyrmion,
but also noticeably improve agreement with phenomenology.}

On an intuitive level, to see the relation between the 
instanton and the Skyrmion,
one can consider as an example the well-known instanton solution in the
singular gauge (in the flat background metric):
\begin{equation}
A_\mu =   \frac{\tau^a \bar\eta^a_{\mu\nu}x_\nu}{x^2(x^2+\rho^2)}\,.
\end{equation}
In this solution we shall think of $A_\mu$ 
as a four-vector with coordinates $\mu$ running through 1, 2, 3, and 5
(i.e., $x$, $y$, $z$, and $u$) and $x^2=\bm x^2 + u^2$. Note
that the metric signature for these 4 coordinates is Euclidean
$(-,-,-,-)$ (see \eq{metric}).
The solution we wish to use to describe a baryon
is static, i.e., $A_0=0$, and there is no dependence on
$t$. To see the behavior of the pion field we need to look at
$A_5$ (see \eq{Sigmacont}):
\begin{equation}
A_5 =   \frac{\bm{\tau \cdot x}}{x^2(x^2+\rho^2)}\,.
\end{equation}
%
We see that at every fixed $u$ the solution is a hedgehog,
thus having the same topology as the Skyrmion made of the
pion field.

We shall now show that for an arbitrary background metric
the topological charge of the instanton is equal to the
baryon charge of the pion-field Skyrmion.  
Our discussion is very similar to
that of Refs.~\cite{HillRamond,Hill}.
The 5d Yang-Mills theory possesses a conserved topological current,
\begin{equation}\label{5dtopcurr}
  \sqrt{\dg}\, j_{\rm 5d}^{\hat\mu} = \frac1{32\pi^2} 
  \emnlrs
  \Tr F_{{\hat\nu}{\hat\lambda}} F_{{\hat\rho}{\hat\sigma}}\,.
\end{equation}
Here
$\emnlrs$ is
defined so that its elements are $\pm1$.  For simplicity we assume that
$g_5$ is a constant and absorb it into the gauge field, and drop the
hat in the 5d Lorentz indices in subsequent formulas.  The boundaries
are assumed to be $u=\pm\infty$.  That this current is conserved,
$\d_\mu(\sqrt{\dg}\,j^\mu)=0$, can be shown by using the Bianchi identity
$D_{[\mu}F_{\nu\lambda]}=0$.  The topological charge of a static
solution is
\begin{equation}\label{QFF}
  Q = \int\!du\,d^3x\, \sqrt {\dg}\, j_{\rm 5d}^0 =
  \frac1{32\pi^2}\int\!du\,d^3x\,
  \epsilon^{0\mu\nu\lambda\rho}\Tr F_{\mu\nu} F_{\lambda\rho}\,.
\end{equation}
The numerical coefficient in Eq.~(\ref{5dtopcurr}) was chosen so that
the static instanton has unit total charge.

Now consider a field configuration where the pion field, given by the
Wilson line along the $u$ coordinate \eq{Sigmacont} has a nontrivial
winding, and $A_{\mu}$ goes to 0 at the boundaries. 
To compute the
topological charge of this configuration, it is convenient to perform
a gauge transformation to set $A_5=0$.  Explicitly,
\begin{equation}
  A_\mu \to U A_\mu U^{-1} + iU\d_\mu U^{-1}
  \,, \qquad
  U(u,\bm x) = P\exp\Bigl(-i\int_{-u_0}^u\!du'\, A_5(u',\bm x)\Bigr). 
\end{equation}
According to \eq{Sigmacont}, 
$U(+u_0,\bm x)=\Sigma^{-1}(\bm x)$. Thus
while $A_i$ remains 0 at the left boundary $u=-u_0$, it becomes
nonzero at the right boundary:
\begin{equation}\label{Aibound}
  A_i = i\Sigma^{-1}\d_i\Sigma\,,\qquad u\to+u_0\,,\qquad i=1,\,2,\,3.
\end{equation}
By using the identity
\begin{equation}
  \epsilon^{0\mu\nu\lambda\rho} \Tr F_{\mu\nu} F_{\lambda\rho} =
  \d_\mu K^{0\mu}\,,\qquad
  K^{0\mu} = 4\epsilon^{0\mu\nu\lambda\rho}
  \Tr\Bigl(A_\nu\d_\lambda A_\rho - \frac{2i}3 A_\nu A_\lambda A_\rho\Bigr),
\end{equation}
one can rewrite the topological charge (\ref{QFF}) as
\begin{equation}
  Q = \frac1{32\pi^2}\int\!d^3x\, K^{05}\big|_{u=-u_0}^{u=+u_0} 
    = \frac1{32\pi^2}\int\!d^3x\, K^{05}\big|_{u=+u_0}\,,
\end{equation}
since at $u=-u_0$ $A_i=0$ and $K^{05}=0$.  By using
Eq.~(\ref{Aibound}) one transforms this expression to
\begin{equation}
  Q = \frac i{24\pi^2}\int\!d^3x\, \epsilon^{ijk}
  \Tr[(\Sigma^{-1}\d_i\Sigma)(\Sigma^{-1}\d_j\Sigma)
  (\Sigma^{-1}\d_k\Sigma)].
\end{equation}
It is now obvious that the topological charge becomes the winding
number of the pion field.  Therefore, the instanton becomes a
Skyrmion, and corresponds to the physical baryon.

\section{Conclusions and discussion}
\label{sec:concl}

We considered a theory of an ``open moose'' given by Lagrangian \eq{L}
illustrated in Fig.~\ref{fig:K}.  This model describes 
a multiplet of massless Goldstone
bosons and a tower of vector and axial
vector mesons. We developed a formalism for calculating
the mass spectrum and the coupling constants in this theory for
arbitrary parameters of the moose, $f_k$ and $g_k$,
and determine their values in the continuum limit, when
the number of hidden symmetry groups $K$ tends
to infinity.  We applied this formalism to two exactly
solvable realizations of the model and found that the physics of the
lowest modes match quite well with the phenomenology of the $\pi$, $\rho$
and $a_1$ mesons.

We also find that the open-moose theory naturally incorporates
the phenomenon of vector meson dominance.
For example, the pion
form factor is saturated by poles from
a tower of vector mesons.  Moreover,
since couplings between mesons are given by
overlap integrals, the couplings of highly excited $\rho$'s to the pion
are suppressed by the oscillations of their wave functions in the fifth
dimension.  This means that the pion form factor should be
well approximated by the
sum of contributions from a few lowest $\rho$'s.  
In the second example we considered (the ``cosh''
background) the situation is brought to an extreme: the pion form factor is
saturated by a single pole -- $\rho$-meson dominance.
We verified that both Weinberg's spectral sum rules are automatically
obeyed, in a nontrivial way, in any open-moose theory.

One of our original motivations was to include the excited vector
mesons beyond the lowest $a_1$.  With respect to that goal, we achieved only
limited success, at least within the two exactly solvable models we
considered.  On the one hand, we do find that vector and axial
vector mesons alternate in the spectrum, as it seems to be the case in
QCD, at least for a few excited states.  On the other hand, in both our
simple models, the mass of an $n$-th state $m_n$ is ${\cal O}(n)$ 
at large $n$, 
which seems to be in contradiction with the real world, and with the
theoretical prejudice that $m_n={\cal O}(\sqrt n)$. Further study of
different backgrounds might provide a model that reproduces desired
features of excited mesons and help understand constraints that
phenomenology and
QCD theorems impose on functions $f(u)$ and $g(u)$.
Alternatively, it is also possible that the excited vector mesons 
have ``stringy'' nature and
cannot, in principle, be incorporated into our field-theoretical scheme.%
\footnote{It is possible to reproduce the behavior 
$m_n={\cal O}(\sqrt n)$ 
by a suitable choice of background, even an exactly solvable one.
But we did not find such models viable in other respects.}

The success that the model enjoys in describing the lowest
states can be attributed to an 
apparent property of low-energy QCD: at
intermediate distances correlation functions are reasonably well
saturated by a single pole.  In the ``cosh'' model the excited mesons
ensure the correct behavior of the (averaged) spectral densities, thus
playing the role of the continuum.  This explains why some
results of QCD sum rules are well reproduced.

We hope that the study of the open moose theories will deepen
our understanding of QCD at the fundamental level.  One intriguing
fact discovered in these theories is the
similarity to the AdS/CFT correspondence.
The procedure of calculating current-current correlators is
essentially equivalent to the well-known AdS/CFT prescription:
the correlators are given by the variational derivatives
of the classical 5d action of the dual theory with respect
to the sources living on the 4d boundary.
There is overwhelming evidence that the
${\cal N}=4$ supersymmetric Yang-Mills theory is described by a string theory.
Perhaps, an open moose theory is a low-energy limit of the string
theory dual to QCD.\footnote{From this point of view, 
meson interactions in strongly coupled gauge 
theories with fundamental quarks~\cite{KarchKatz,KruczenskiMyers}
deserve further studies.}
 In this regard, the result we found
in the ``cosh'' model, 
\begin{equation}
g_5^2\sim \frac1{N_c},\nonumber
\end{equation}
is reassuring in view of
a general expectation that such a dual theory should
have a coupling proportional to $1/N_c$ in the 't Hooft limit.

Among the questions left for further study is the detailed phenomenology of
isoscalar mesons ($\eta$, $\omega$, $f_1$, etc.).
These mesons are described by an additional
 5d Abelian gauge field, which should be introduced into the
action~(\ref{5dS}). Most of our results should generalize straighforwardly
to this case. However, there is an important new issue that the
isoscalar sector brings into the theory. The global 
U(1)$_A$ symmetry must be explicitly broken, e.g., $\eta$ should not
be massless. It is very encouraging that the 5d formulation
of the theory provides a very natural mechanism for this.
It is the topological 5d Chern-Simons term of the form 
\begin{equation}\lb{AFF}
\int\! d^5x\, \emnlrs
A_{\hat\mu} F^a_{\hat\nu\hat\lambda} F^a_{\hat\rho\hat\sigma},
\end{equation}
where $A_{\hat\mu}$ is the 5d vector field describing isoscalars.
This term breaks the U(1)$_A$ symmetry in the desired way. In particular,
it is not invariant under U(1)$_A$ transformations
on the 4d boundary (although it is invariant under local
transformations in the bulk of 5d).
It is easy to see that it also provides $\pi^0\to2\gamma$ and
 other anomalous processes in QCD. The coefficient of
the term \eq{AFF}
can be fixed by matching to QCD chiral anomaly, and is therefore
proportional to $N_c$.  The term \eq{AFF} also 
couples the $\omega$ meson field to the baryon current, providing a
hard-core repulsion between baryons, and preventing the
baryon/instanton from shrinking to zero size (this effect is the
origin of the stabilization of the Skyrmion observed in
Ref.~\cite{AdkinsNappi}). It would be also interesting to
see how the open-moose theory
realizes Di~Vecchia-Veneziano-Witten Lagrangian~\cite{DiVV}
and the corresponding phenomenology.
Other avenues for future
study are the incorporation of finite quark masses,
extension to three flavors and realization of the
Wess-Zumino-Witten topological term (which does require a 5th 
dimension~\cite{Witten-WZW}).

{\bf Acknowledgments}

The authors thank S.~R.~Beane, G.~Gabadadze,
W-Y.~Keung, and especially T.~Imbo and M.~Strassler
for discussions.   The authors also wish to acknowledge the review
talk by R.~L.~Jaffe at the {\em QCD and String Theory} workshop at the
Institute for Nuclear Theory in 
Seattle, which provided much of the motivation for this work.
M.A.S. acknowledges the hospitality of
the Institute for Nuclear Theory, University of Washington,
where part of this work has been done, and thanks RIKEN
BNL Center and U.S. Department of Energy [DE-AC02-98CH10886] for
providing facilities essential for the completion of this work.
D.T.S. is supported, in part, by DOE grant No.\
DOE-ER-41132 and the Alfred P.\ Sloan Foundation.  M.A.S. is
supported, in part, by a DOE OJI grant and by the Alfred P.\ Sloan
Foundation.

\appendix

\section{Interaction vertices}
\lb{app:mnp}

For reference, we provide here some additional formulas for
interaction vertices in the continuum limit of an arbitrary
open-moose  model.
Let us define $g_{\pi mn}$, $g_{\pi\pi mn}$,
$g_{mnp}$, and $g_{mnpq}$ so that the Lagrangian contains
\begin{equation}\lb{couplings}
\begin{split}
  {\cal L} &= \cdots 
  - g_{\pi mn}\epsilon^{abc}\pi^a (\alpha^{m}_\mu)^b (\alpha^{n}_\mu)^c
  -g_{\pi\pi mn}\epsilon^{abc}\epsilon^{ade}
  \pi^b\pi^d(\alpha_\mu^{m})^c(\alpha_\nu^{n})^e\\
  &\quad- 
  g_{mnp}\epsilon^{abc}(\alpha^{m}_\mu)^a(\alpha^{n}_\nu)^b
\d_\mu(\alpha^{p}_\nu)^c
  - \frac14 g_{mnpq}\epsilon^{abc}\epsilon^{ade}
  (\alpha^{m}_\mu)^b (\alpha^{n}_\nu)^c (\alpha^{p}_\mu)^d (\alpha^{q}_\nu)^e,
\end{split}
\end{equation}
then
\begin{subequations}
\begin{eqnarray}
  g_{\pi mn} &=& \frac{f_\pi}4 \int\!du\,
      [(gb_m)(gb_n)'-(gb_m)'(gb_n)],\lb{pimn}\\
  g_{\pi\pi mn} &=& -\frac{f_\pi^2}8\int\!\frac{du}{f^2(u)}\, 
      b_m b_n,\lb{pipimn}\\
  g_{mnp} &=& \int\!du\, gb_m b_n b_p,\lb{mnp}\\
  g_{mnpq} &=& \int\!du\, g^2 b_m b_n b_p b_q,\lb{mnpq}
\end{eqnarray}
\end{subequations}
Direct couplings to external currents are suppressed in the
continuum limit (this includes, in particular, vector-meson dominance
by a whole tower of mesons).

A simple qualitative interpretation of these couplings exists in terms of the
overlaps of the wave functions in the $u$ space, which reflects
the property of locality of the theory \eq{L} or \eq{5dS}. 
It is straightforward
for the last two, resonance-resonance, couplings \eq{mnp} and
\eq{mnpq}. These terms come from the second, $F_{\mu\nu}^2$ term in \eq{L}.
For the pion-resonance couplings \eq{pimn}, \eq{pipimn} and \eq{gnpipiu},
one should bear in mind that the strength of the coupling is
proportional to $f^2(u)$, and think of the pion wave function
as being proportional to $1/f^2$ (looking at \eq{SigmaPi}).
The $u$ derivatives in \eq{pimn} are necessary to account for the
fact that, although the pion wave function is even in $u\to-u$,
the pion itself is a pseudoscalar.

\section{Summary of results for the ``cosh'' model}
\label{app:cosh}

Instead of expressing the results in terms of the parameters of the
model $\Lambda$ and $g_5$, we will use $m_\rho$ and $N_c$.  The
relations are
\begin{equation}
  m_\rho = \sqrt{2}\Lambda\,, \qquad g_5^2 = \frac{24\pi^2}{N_c}\,,
\end{equation}
\begin{equation}
  m_n = m_\rho\sqrt{\frac{n(n+1)}2}\,,
\end{equation}
\begin{eqnarray}
  f_\pi = \frac{m_\rho}{2\pi}\sqrt{\frac{N_c}6}\,,
\end{eqnarray}
\begin{eqnarray}
  g_{nV,A} = \frac{m_\rho^2}{4\pi}\sqrt{\frac{n(n+1)(2n+1)}3 N_c}\,,
\end{eqnarray}
Rho-meson dominance of the pion form factor:
\begin{equation}
 g_{n\pi\pi}=0 \,, \qquad n\ne1,
\end{equation}
\begin{equation}
 g_{1\pi\pi}\equiv g_{\rho\pi\pi} = \frac{2\sqrt2\pi}{\sqrt{N_c}}
\quad
\Bigl(\mbox{also}\quad g_{\rho\pi\pi}=\frac{m_\rho^2}{g_{\rho V}}\Bigr).
\end{equation}
There is a ``$\Delta n =1$ rule'' for pion emission:
\begin{equation}
  g_{\pi mn} = 0\,, \qquad |m-n|\ne1,
\end{equation}
\begin{equation}\lb{gnn+1}
  g_{\pi\, n\, n+1} = m_\rho \pi(n+1)\sqrt{\frac{6n(n+2)}
  {(2n+1)(2n+3)N_c}}\,.
\end{equation}
There is also a ``triangle rule'' for triple resonance vertex:
\begin{equation}
\begin{split}
  g_{mnp} = 0 \quad \textrm{if} \quad & m+1>n+p+2\\ 
                    & \textrm{or $n+1>p+m+2$}\\ 
                    & \textrm{or $p+1>m+n+2$},
\end{split}
\end{equation}
i.e., the amplitude vanishes if a triangle (even a degenerate one) 
with sides $(m+1)$, $(n+1)$ and $(p+1)$ does not exist.

\section{$a_1$ meson in the ``cosh'' background}
\lb{app:a1}

Let us discuss the phenomenology of the lowest axial vector meson (the
$n=2$ excitation in the open moose).  
From Eq.~\eq{mncosh}, the mass of the $a_1$ meson in
the ``cosh''  model is $m_{a_1}=\sqrt3m_\rho$.  The $a_1$ decays 
into $\rho\pi$
with the coupling \eq{gnn+1}
\begin{equation}
  g_{\pi\rho a_1} = 2\pi m_\rho \sqrt{\frac6{5N_c}}\,.
\end{equation}
By using the formula~\cite{HLSa1}
\begin{equation}
  \Gamma(a_1\to\rho\pi) = \frac{g_{\pi\rho a_1}^2}{4\pi} p_\rho
  \Bigl(1+\frac{p_\rho^2}{3m_\rho^2}\Bigr),
\end{equation}
we find
\begin{equation}
  \Gamma(a_1\to\rho\pi) = \frac{4\pi}{9\sqrt{3}N_c} m_\rho \approx 210~
   \textrm{MeV}.
\end{equation}
Experimentally, the total width of $a_1$ is 250 to 600 MeV, of which
about 60\% comes from $a_1\to\rho\pi$~\cite{PDG}.

The $a_1$ decay constant in our model is
\begin{equation}
  g_{a_1A} \equiv g_{2A} = \frac{m_\rho^2}{2\pi}\sqrt{\frac{5N_c}2}
  \approx 0.26~\textrm{GeV}^2.
\end{equation}
A lattice measurement of this constant yields (in our normalization)
$0.21\pm0.02$ GeV$^2$~\cite{Wingate}, while an analysis of
hadronic $\tau$ decays gives $0.177\pm0.014$
GeV$^2$~\cite{Isgur}.  The agreement is fair, but not exceptionally
good.

\begin{figure}[htbp]
  \begin{center}
    \epsfig{file=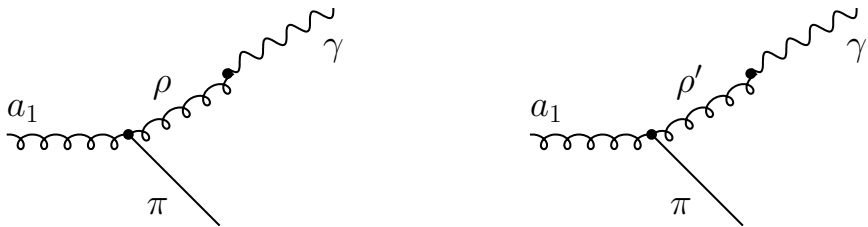,width=.7\textwidth}
  \end{center}
\caption[]{Diagrams contributing to $a_1\to\pi\gamma$.}
    \label{fig:a1pigamma}
\end{figure}

The decay $a_1\to\pi\gamma$ occurs through two Feynman diagrams with
intermediate $\rho$ and $\rho'$ (the $n=3$ excitation).  Its amplitude is
proportional to (see Fig.~\ref{fig:a1pigamma})
\begin{equation}
  \frac{g_{\pi1 2}{g_{1 V}}}{m_1^2} +
  \frac{g_{\pi3 2}{g_{3 V}}}{m_3^2}
\equiv
  \frac{g_{\pi\rho a_1}{g_{\rho V}}}{m_\rho^2} +
  \frac{g_{\pi\rho' a_1}{g_{\rho' V}}}{m_{\rho'}^2}\,.
\end{equation}

It can be checked that the two terms cancel each other exactly, so the
amplitude vanishes.  On the other hand, the partial width of this
decay is quoted to be $640\pm246$ keV~\cite{PDG}.  The simplest $K=2$
hidden local symmetry model also suffers from the same problem; in
Ref.~\cite{HLSa1} this was cured by adding higher-derivative terms to
the action.  It would be interesting to see if this rate can be made
nonzero by adding more terms to the 
action~(\ref{5dS}).

\end{document}